\documentclass[12pt,preprint]{aastex}

\usepackage{graphicx}
\usepackage{dcolumn}
\usepackage{bm}

\begin{document}
\title{Particle acceleration in the driven relativistic reconnection}
\author{Yuri Lyubarsky, Michael Liverts}
\affil{Department of Physics, Ben Gurion University of the
Negev, Beer-Sheva, Israel}

\begin{abstract}
We study the compression driven magnetic reconnection in the
relativistic electron-positron plasma. Making use of a 2.5D
particle-in-cell code, we simulated compression of a magnetized
plasma layer containing a current sheet within it. We found
that the particle spectrum within the reconnecting sheet
becomes non-thermal; it could be approximated by a power-law
distribution with an index of -1 and an exponential cutoff.
\end{abstract}
\keywords{plasmas, magnetic fields, acceleration of particles}
\section{Introduction}
Magnetic reconnection is one of the important processes for the
particle acceleration and heating. The production of energetic
particles has been extensively studied in non-relativistic
reconnection \cite{Cargill01, Drake05, Pritchett06}, mostly in
the solar terrestrial context. Magnetic reconnection may also
play a major role in relativistic objects such as pulsars
\cite{Kirk07}, magnetars \cite{Thompson95,Lyutikov03}, active
galactic nuclei \cite{Romanova92,diMatteo98,Birk01} or
gamma-ray bursts
\cite{Drenkhahn02,Drenkhahn-Spruit02,Thompson06}. Genuinely
relativistic reconnection occurs if the magnetic energy exceeds
the plasma energy including the rest energy. Then dissipation
of the magnetic field inevitably leads to relativistic energies
of the particles \cite{Kirk04}.

Relativistic reconnection typically occurs in electron-positron
plasmas; in this case physics of the process is simplified
considerably as compared with electron-ion plasmas where two
significantly different scales are presented. Larrabee, Lovelace
\& Romanova (2003) calculated self-consistent equilibria of the
relativistic electron-positron plasma in the vicinity of the
magnetic X-point in the reconnecting current sheet. The particles
are accelerated there by the electric field parallel to the
X-line. It was found that the particle distribution function is
described by a power law $dn/d\gamma\propto \gamma^{-1}$ with an
exponential cutoff. Note that generally only a small fraction of
magnetic energy is released in the vicinity of the X-point.
Reconnection in the X-point just allows the magnetic field lines
to shrink thus releasing the magnetic energy. For example, in the
classical Petschek model, most of the energy is released at slow
shocks that stem from the X-point. Therefore in order to find
overall particle distribution function, one have to study particle
acceleration/heating in the whole reconnection region; the results
will presumably depend on global geometry and boundary conditions.
Particle-in-cell (PIC) simulations of spontaneous reconnection in
an infinite, plane current sheet (Zenitani \& Hoshino 2001, 2005,
2007, 2008; Jaroschek et al. 2004) confirm that a dc acceleration takes
place around the X-point and that the particle energy spectrum in
this region is roughly described by a power law with the power
index of -1. However, the energy spectrum over the whole
simulation domain is significantly steeper; it is approximated by
the power law with an index of -3 \cite{Jaroschek04,Zenitani07}.

This paper is aimed at the particle acceleration in the
compression driven reconnection. Our study is motivated by
observation (Lyubarsky 2003, 2005; P\'{e}tri \& Lyubarsky 2007)
that reconnection could be an essential part of dissipation
mechanism at the termination shock in the striped pulsar wind.
Pulsars lose their rotational energy predominantly on
generation of the Poynting dominated winds. Most of the energy
is transferred in the equatorial belt where the sign of the
magnetic field alternates with the pulsar period forming
stripes of opposite magnetic polarity; such a structure is
called a striped wind (e.g., review by Kirk et al. 2007). When
the striped wind arrives at the termination shock, the plasma
is strongly compressed; in the comoving frame, the compression
ratio is very large, about the Lorentz factor of the upstream
flow. Therefore the alternating magnetic fields are easily
annihilated by the compression driven reconnection. This
conjecture is supported by 1.5D PIC simulations
\cite{Lyubarsky05,Petri07}.

In one-dimensional simulations, the magnetic energy is
transformed into heat because all the particles gain energy
with the same rate. In order to study particle acceleration,
multidimensional simulations are necessary because one can
expect a formation of non-thermal tails in particle spectrum if
different particles gain different energy; this could happen if
X-points are formed within the current sheet. As a step towards
the multidimensional simulations of the shock in a striped
wind, we performed 2.5D PIC simulations of driven magnetic
annihilation within a plasma layer containing only two stripes
of the opposite magnetic polarity. In our simulations, the
plasma layer is compressed by an external force, which imitates
the compression within the shock structure. We show that within
such a structure, a non-thermal particle spectrum is formed,
which can be roughly described by a power law with the slope -1
and an exponential cutoff. We analyze the particle motion
within the compressed layer and show that the particle
acceleration in the vicinity of X-points plays a crucial role
even though only a small fraction of the energy is released
there. The reason is that in the compressing medium, the larger
the particle Larmor radius, the more energy it gains therefore
particles preaccelerated in the X-point gain more energy than
the particles bypassed the X-point. Thus a large fraction of
the total energy is eventually transferred to particles that
passed the X-point.

The paper is organized as follows. In Section 2, simulation
parameters and methods are introduced. In Section 3, the results
are shown and discussed. Section 4 gives the conclusions of this
work.

\section{Simulation setup}

We used 2.5D (2D3V - 2 spatial and 3 velocity components) fully
relativistic electro-magnetic particle-in-cell code. The
evolution of electric and magnetic fields is governed by
Maxwell equations for components of the field vectors $E_x$,
$B_y$ and $B_z$. In two-dimensional configuration with $B_x=0$,
the components $E_y$ and $E_z$ of the electric field are
decoupled from the remaining field components and could arise
only due to charge fluctuations (cf. simulations by Karlicky
(2007) and Zenitani \& Hoshino (2008) where the charge separation
arises at $B_x \ne 0$). In electron-positron plasma with the
same distribution functions for both species, such fluctuations
are very low and therefore we take $E_y=E_z=0$ throughout the
simulations. Such a suppression of the electrostatic
fluctuations permits the use of relatively small number of
particles.

The fields are updated using the leap-frog scheme. The
staggered grid mesh system, known as Yee lattice \cite{Yee66},
ensures that the change of the magnetic flux through a cell
surface equals the negative circulation of the electric field
around that surface and the change of the electric flux through
a cell surface equals the circulation of the magnetic field
around that surface minus the current through it. Then the
divergency-free condition is maintained to the machine
accuracy. Here the electric and the magnetic fields are in a
symmetry form except subtracting the charge flux $J$ in Ampere
equation. The current density is calculated and subtracted
after the particles are moved later in the program
\cite{Buneman93}. Before and after moving (or pushing) the
particles, the magnetic field is updated in two half steps so
that it is available at the same time as the electric field for
the particle update. The particles positions and velocities are
advanced by Newton-Lorentz 3D equations of motion, which were
solved by the Buneman-Boris method \cite{Birdsall85}.

\begin{figure}[h]
\begin{center}
\includegraphics[width=.45\columnwidth]{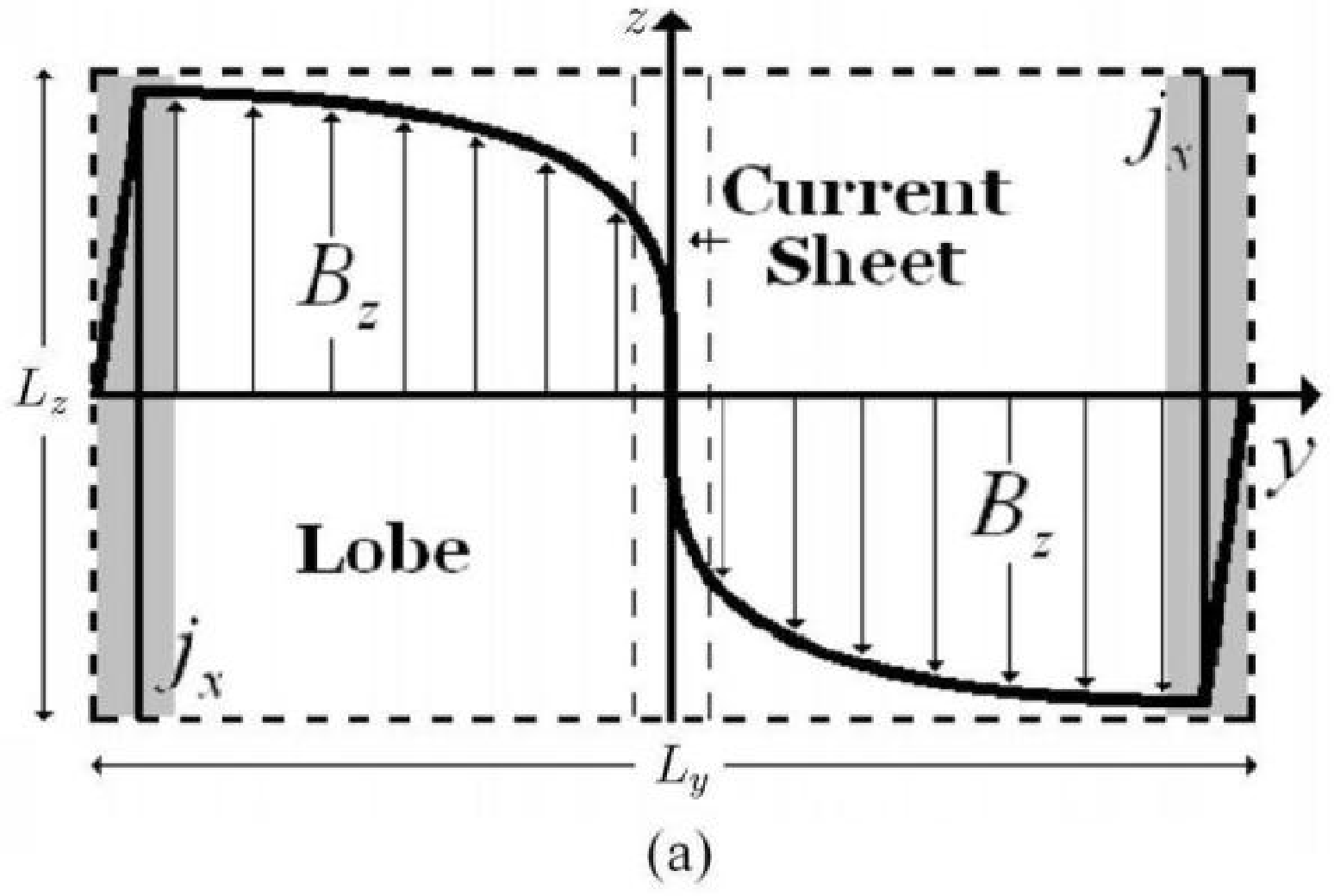}
\includegraphics[width=.46\columnwidth]{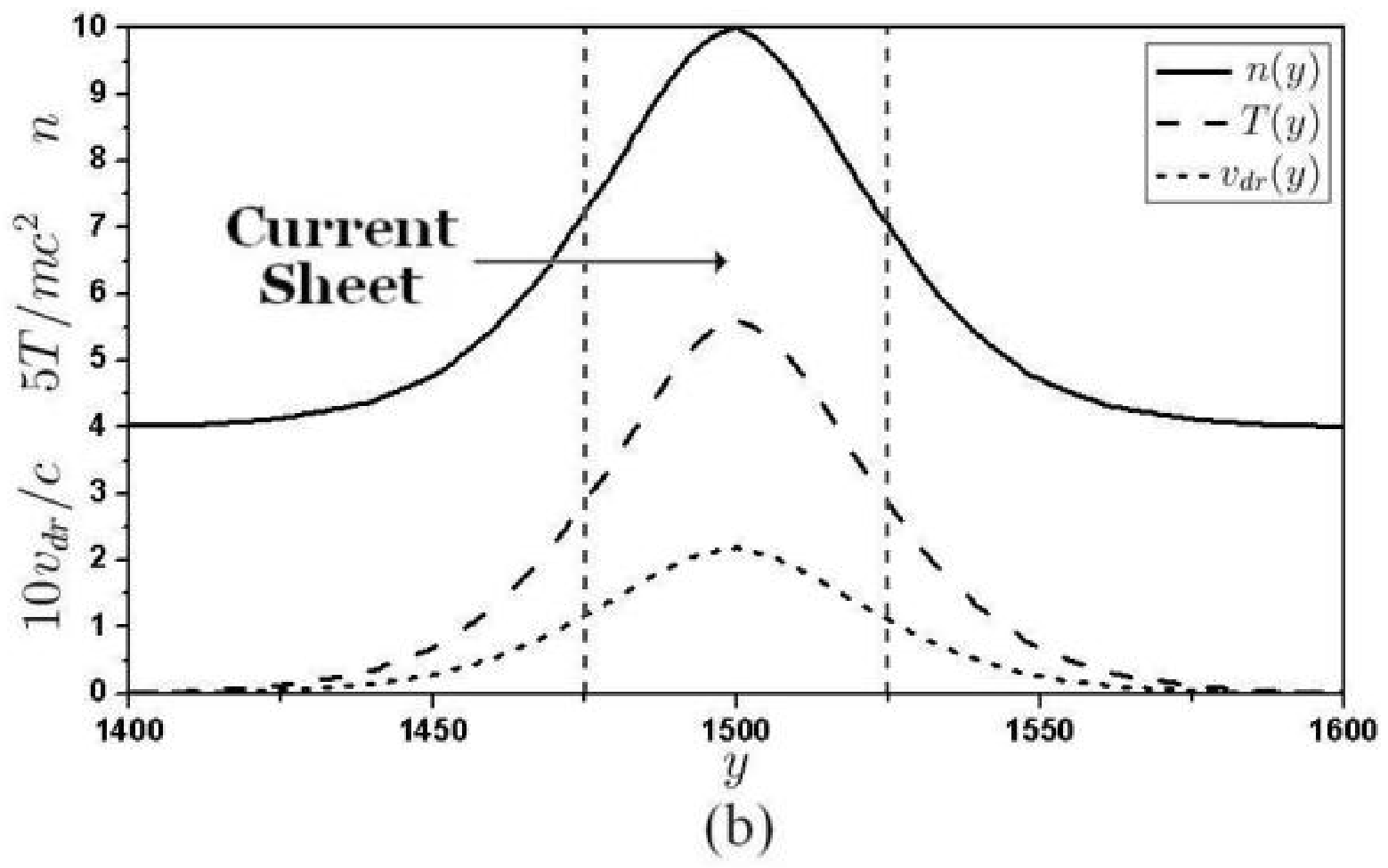}
\end{center}
\caption{(a) Schematics of initial configuration, fields and charge fluxes.
(b) Initial profiles of density $n(y)$, temperature $T(y)$ and velocity $v_{dr}(y)$ near the current sheet.
Outside the central part of the simulation box, the density
remains constant up to the box margins $0<y<10$ and
$2990<3000$, where the plasma is absent ($n=0$)} \label{scheme}
\end{figure}

Assuming the reconnecting current sheet to lie in the $y$-$z$
plane, Fig.\ref{scheme}a illustrates the configuration of the
simulated domain (rectangular grid shown by dashed line). The
system lengths in the $y$- and $z$-directions are $L_y$ and
$L_z$ respectively. The domain is divided into
$3000(L_y)\times400(L_z)$ cells grid or normalized to electron
skin depth $384(L_y \omega_p/c)\times 51(L_z \omega_p/c)$,
where the plasma frequency $\omega_p$ is calculated for the
particle density in the lobe. Fig.\ref{scheme}b demonstrates
the initial profiles of density $n(y)$, temperature $T(y)$ and
velocity $v_{dr}(y)$ near the current sheet ($y=1400-1600$).
Outside the central part of the simulation box, the density
remains constant up to the box margins $0<y<10$ and
$2990<3000$.

To avoid escaping in the $y$-$z$ plane, the particles are
seeded away from boundaries at a distance larger than particles
gyroradius. Ten mesh points are set to keep the margin on each
side of the boundaries (shown gray), i.e., particles are loaded
in the region of $10<y<2990$. At the boundaries, two external
currents normal to the simulation plane are set (as illustrated
in Fig.\ref{scheme} by solid lines at gray margins) in the same
directions. The magnetic field is directed along $z$; in the
region between the currents it is set initially as
\begin{equation}
B_z(y)=B_0\tanh\frac{L_y-2y}{2\Delta}; \label{B}
\end{equation}
where $\Delta(=25)$ is the sheet halfwidth. The initial
magnitude of the external currents is
\begin{equation}
J_0=\frac{cB_0}{q}
\end{equation}
so that the magnetic field vanishes in the region outside the
currents on both sides.

The magnetic field (\ref{B}) implies the current in the
$x$-direction; the current density is determined by Ampere's
law
\begin{equation}
\nabla \times \mathbf{B}=\frac{4\pi \mathbf{J}}{c}.
\end{equation}
This current is generated by motion of electrons and positrons in
opposite directions with the velocity
\begin{equation}
v_{dr}(y)=\frac{c}{2qn(y)}\frac{dB_z(y)}{dy}.
\end{equation}
The initial spatial distribution of particles is non-uniform
along $y$-direction
\begin{equation}
n(y)=n_0+\frac{n_{max}-n_0}{\cosh((L_y/2-y)/\Delta)}.
\end{equation}
In the main simulations presented here, the total number of
particles is $5\cdot10^6$, giving $n_0=4$ particles per cell in
the lobe and $n_{max}=10$ in the sheet. In order to check
reliability of our simulations, we have also run similar models
with smaller box $300(L_y)\times40(L_z)$ but with different
number density of particles, where $n$ was multiplied by factor
$\alpha=1\div10$. In order to keep the system in mechanical
equilibrium (\ref{equilibrium}), the magnetic field $B$ was
multiplied by $\sqrt{\alpha}$, while other parameters were kept
unchanged. We found that particles spatial distribution and
energy spectrum during the compression were independent of the
number density $n$.

The particles are initialized by the Maxwell distribution with
a temperature depending on $y$ such that the system is in
mechanical equilibrium, i.e.
\begin{equation}
nT+\frac{B^2}{8\pi}=\mbox{const}. \label{equilibrium}
\end{equation}
Then
\begin{equation}
T(y)=\frac{B_0^2-B_z(y)^2}{16\pi
n(y)\sqrt{1-\left(v_{dr}(y)/c\right)^2}}+T_0
\end{equation}
where $T_0$ is a minimal temperature value, $T_0=0.1m_ec^2$.
The chosen initial state is close to the true kinetic
equilibrium because the initial thickness of the sheet is
significantly larger than the initial particle Larmor radius,
$2\Delta/r_L \approx 29$, where $r_L = \gamma m v c/qB_0
\approx 1.75$ (the thermal velocity $v$ is calculated using the
initial temperature at the center of the sheet, which is
$T=1.22mc^2$).

The imposed boundary conditions along $z$-axis are periodic.
Along $y$-axis, we impose Lindman's radiation-absorbing
condition, which requires that the fields should be able to
radiate away into space and should not be reflected
\cite{Lindman75}. The time step is chosen such that
$\omega_c\Delta t \simeq 0.2$ where $\omega_c=qB_0/mc$. The
plasma in the lobe is magnetically dominated;
$\omega_p/\omega_c$ is initially 0.32 ($n_0$ is used to
calculate $\omega_p$) and decreases in the course of
compression.

\section{Simulation results}
In simulations, the reconnection is driven by compression. We
increase the magnitude of boundary currents on both sides so
that the magnetic field grows and the plasma is compressed
towards the midplane of the simulated domain. Note that in the
course of compression, the plasma layer becomes narrower so
that the pressure balance is restored faster therefore in order
to save the computer time we can take the compression rate
increasing with time. We choose the boundary currents
quadratically growing with time:
\begin{equation}
J(t)=J_0\left[1+\nu\frac{c}{L_y}t\right]^2
\end{equation}
where $\nu=1.3$. We have also run the simulation with the
compression rate $\nu=0.75$ (twice slower) however no
significant difference in the final spatial distribution and
energy spectrum was observed.

\begin{figure}[h]
\begin{center}
\includegraphics[width=.5\columnwidth]{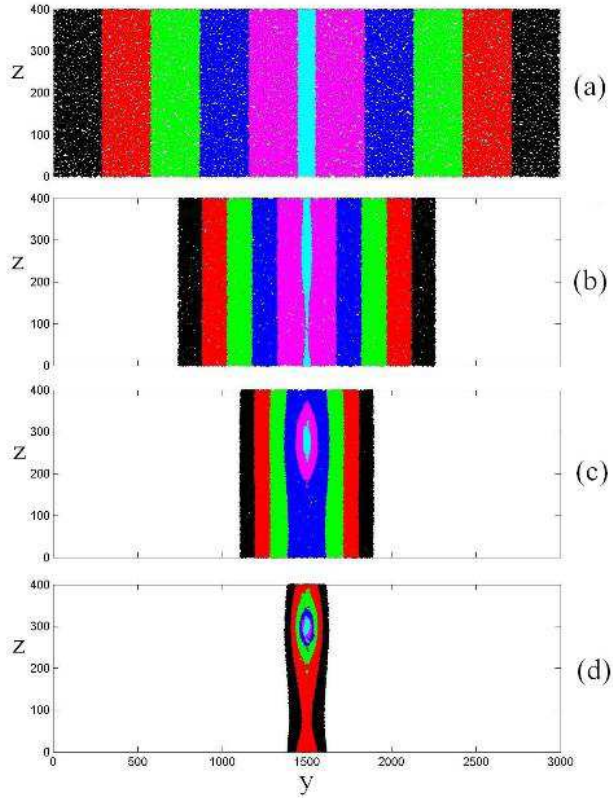}
\end{center}
\caption{Evolution of the particle spatial distribution
initially separated into colored stripes along $y$-direction at
(a) $t=0$, (b) $t=5000$, (c) $t=7000$ and (d) $t=9990$.}
\label{motion}
\end{figure}

As a result of external compression, the magnetic field lines
reconnection occurs in the midplane where the field reverses
sign. Fig.\ref{motion} demonstrates evolution of spatial
distribution of particles.
The simulated domain is initially
separated into several stripes, which are located at different
distances from the midplane. The particles located initially in
each stripe are marked with a corresponding color which allows
tracing the motion of different pieces of plasma in the course
of compression. Fig.\ref{motion}a shows initial distribution of
particles, where due to the reflection symmetry, the stripes at
the same distances from the midplane are colored identically on
both sides. The width of the stripes is chosen the same except
of the central, more dense, stripe (cyan) which is twice as
large as the sheet width. As the boundary currents grow, the
plasma is compressed and each stripe shifts towards the
midplane i.e. to $y=1500$, (see Fig.\ref{motion}b). In the lobe
region, the particles energy is low, therefore the Larmor radii
are small and particles are restricted to move with the
magnetic field lines towards the current sheet. Therefore each
stripe moves as a whole until it reaches the current sheet.
Within the current sheet, the magnetic field lines are
reconnected and as a result particles assemble on the magnetic
island, where each stripe forms a corresponding ring, as one
can see in Fig.\ref{motion}d.

\begin{figure}[h]
\begin{center}
\includegraphics[width=.5\columnwidth]{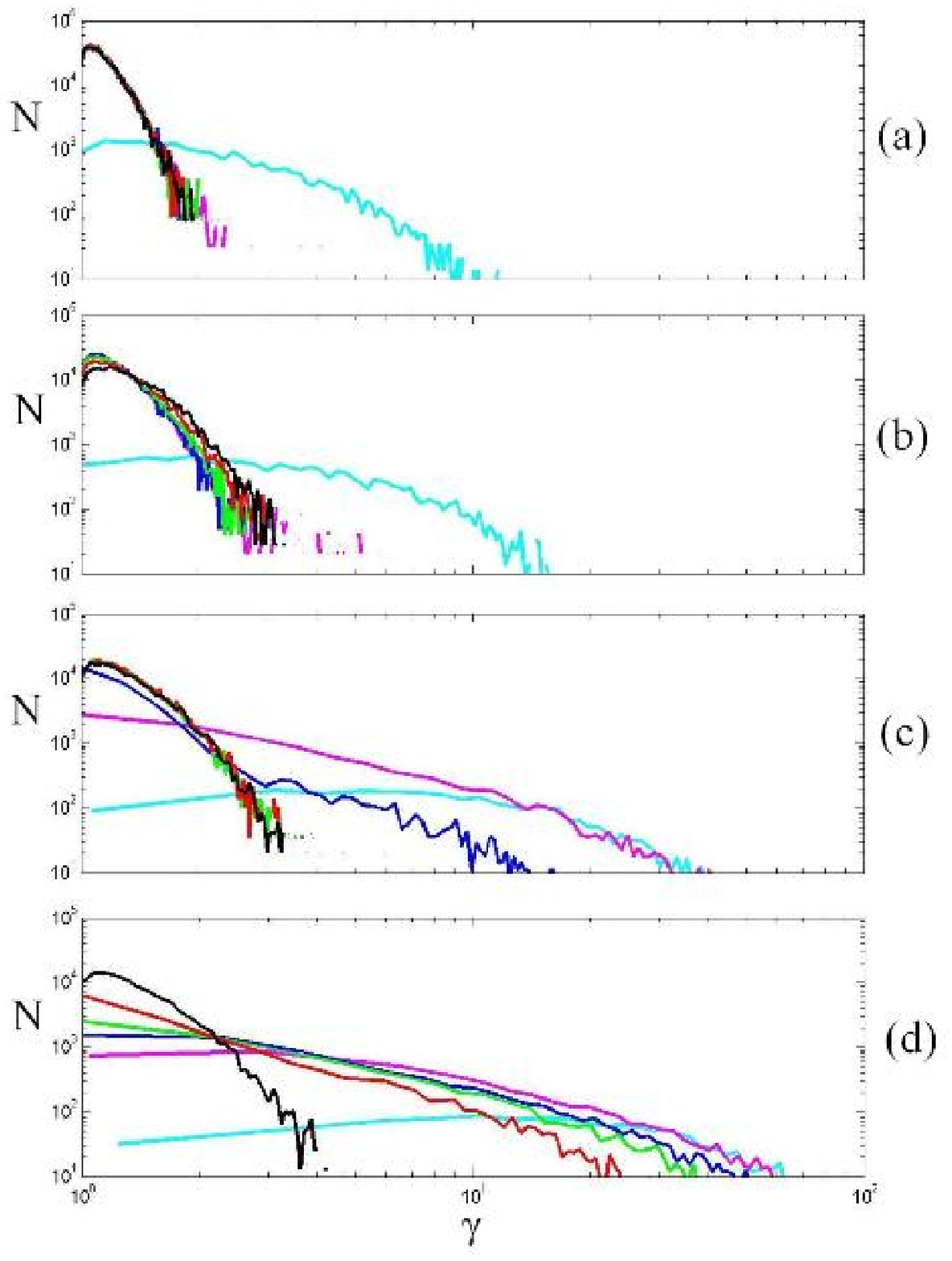}
\end{center}
\caption{Evolution of the particle energy spectrum for each of
the colored stripes at (a) $t=0$, (b) $t=5000$, (c) $t=7000$ and
(d) $t=9990$.} \label{gamma_regions}
\end{figure}

Evolution of the particle energy spectrum in each stripe is
presented at Fig.\ref{gamma_regions}.
Initially, as shown on
Fig.\ref{gamma_regions}a, the energy of particles in the lobe
is low and only in the current sheet, the higher energy
particles are present. When compressed, the stripes begin to
form ring-regions in the sheet. When a stripe assembles in a
ring, the particle energy spectrum extends to higher energies
Fig.\ref{gamma_regions}c shows the spectrum of the magenta
stripe, which begins to form a ring at $t=7000$.
Fig.\ref{gamma_regions}d shows that by the end of simulations,
the hard tail in the energy spectrum is formed predominantly by
particles entered the current sheet in the course of the
reconnection process. The fraction of particles residing in the
sheet from the beginning (cyan curve) is small therefore the
final spectrum is independent of the initial configuration of
the system.

\begin{figure}[h]
\begin{center}
\includegraphics[width=.5\columnwidth]{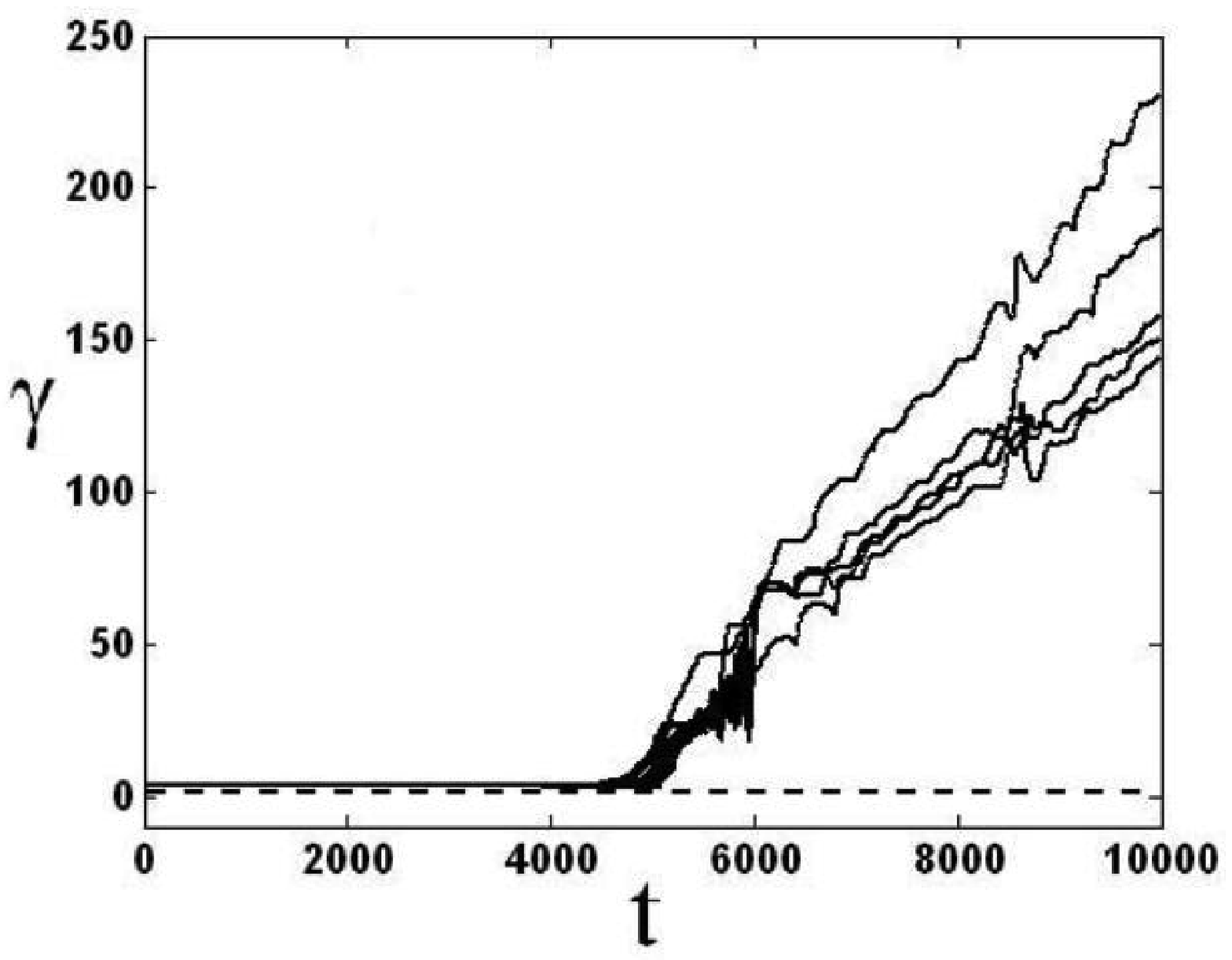}
\end{center}
\caption{Energy of test particles. Solid - high-final energy particles.
Dashed - low-final energy particles.} \label{gamma}
\end{figure}

\begin{figure}[h]
\begin{center}
\includegraphics[width=.5\columnwidth]{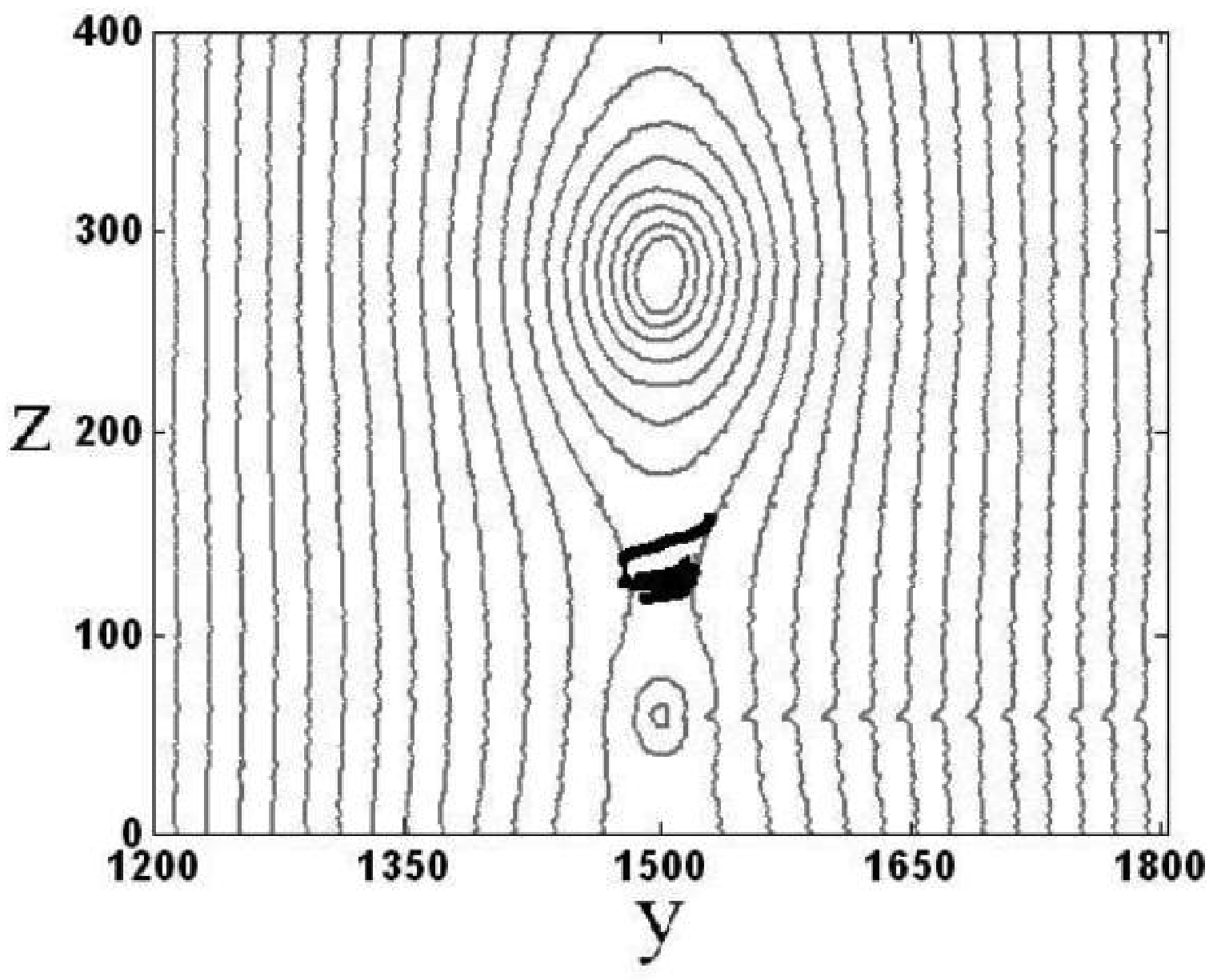}
\end{center}
\caption{Magnetic field lines (gray) and five typical trajectories (black) of high-final energy particles
during the onset of acceleration at $t\approx 5000$ (see Fig.~\ref{gamma}).
All the trajectories (over the interval $\Delta t=100$) are within the X-type region.} \label{B_Traj}
\end{figure}

In the course of compression and magnetic reconnection, all the
particles gain energy. However, taking two identical particles
at the same distance from the sheet center one can find a
significant difference (by two orders of magnitude) in their
final energies. This wide range of final energies demands
further explanation. We take a number of particles with high
and low final energies initially located at the same distance
of 300 from the midplane (i.e. $y=1200$ and $y=1800$) and trace
them during the compression process. Fig.\ref{gamma}
demonstrates the energy evolution of high-final (solid curves)
and low-final (dashed curves) energy test particles.
As shown,
the high-final energy particles begin acceleration at
approximately the same time ($t \approx 5000$) when the chosen
stripe enters the current sheet. The particles from the same
stripe that were not accelerated at this time remain cold. In
Fig.\ref{B_Traj}, five typical trajectories of high-final
energy particles are shown during the short time-interval
($\Delta t=100$) when the particle energy just begins to grow.
The simultaneous magnetic field lines are shown by thin lines.
One sees that, during the onset of acceleration, all these
particles are located in the vicinity of the X-point.
Afterwards the particles leave the X-point however their energy
continues to grow (see Fig.\ref{gamma}), whereas the energy of
particles that bypassed the X-type region remains low. The
reason is that in the course of compression, the particles gain
energy proportionally to the available one. Therefore the
particles preaccelerated near the X-point continue to gain
energy while those that bypassed the X-point remain cold.

\begin{figure}[h]
\begin{center}
\includegraphics[width=.5\columnwidth]{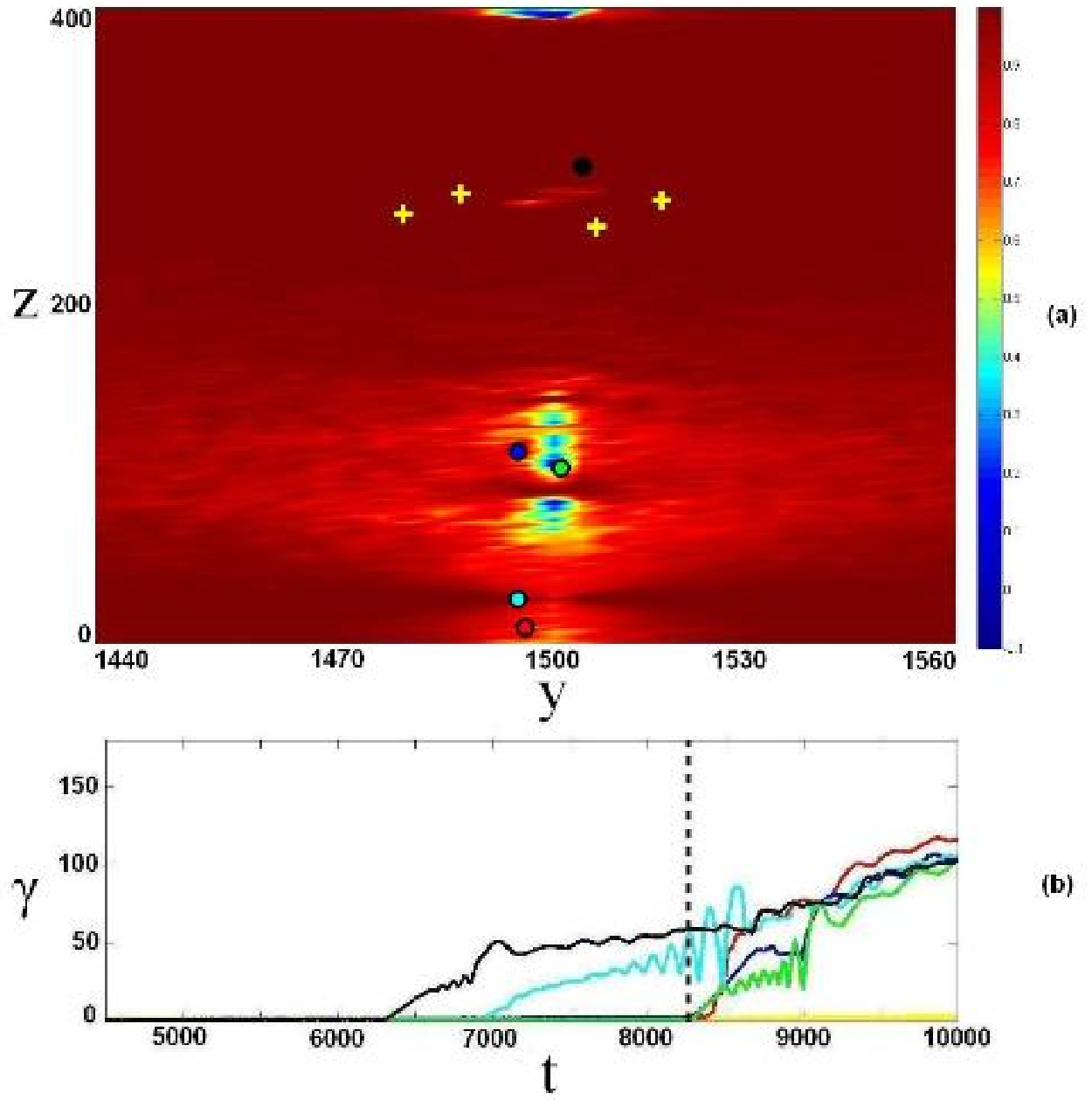}
\end{center}
\caption{(a) Low (yellow crosses) and high (color) energy
particles position over $\chi$-value background at $t=8260$, (b)
Energy evolution of the traced particles.} \label{chi}
\end{figure}

A closer look at the particle trajectory shows that the
steepest jumps in energy appear at certain locations within the
X-point region. The particle energy grows sharply when the
electric field becomes equal or even exceeds the magnetic
field.
In order to trace the acceleration sites, the quantity
\begin{equation}
\chi=\frac{\mathbf{B}^2-\mathbf{E}^2}{\mathbf{B}^2+\mathbf{E}^2}
\end{equation}
was constructed and shown in Fig.\ref{chi}a as a color
background; here $\chi \rightarrow -1$ corresponds to $E \gg B$
and $\chi \rightarrow 1$ corresponds to $E \ll B$.
The
locations of the test particles are also presented for
high-final (colored dots) and low-final (yellow crosses) energy
particles. Fig.\ref{chi}b shows the energy evolution of the
traced particles; the vertical dashed line marks the instant
($t=8260$) of the snapshot at Fig.\ref{chi}a. One can see that
when a test particle (the green and blue dots in
Fig.\ref{chi}a) enters the region of low $\chi$-value, the
energy (the green and blue curves in the bottom panel) of the
particle explosively grows.

\begin{figure}[h]
\begin{center}
\includegraphics[width=.5\columnwidth]{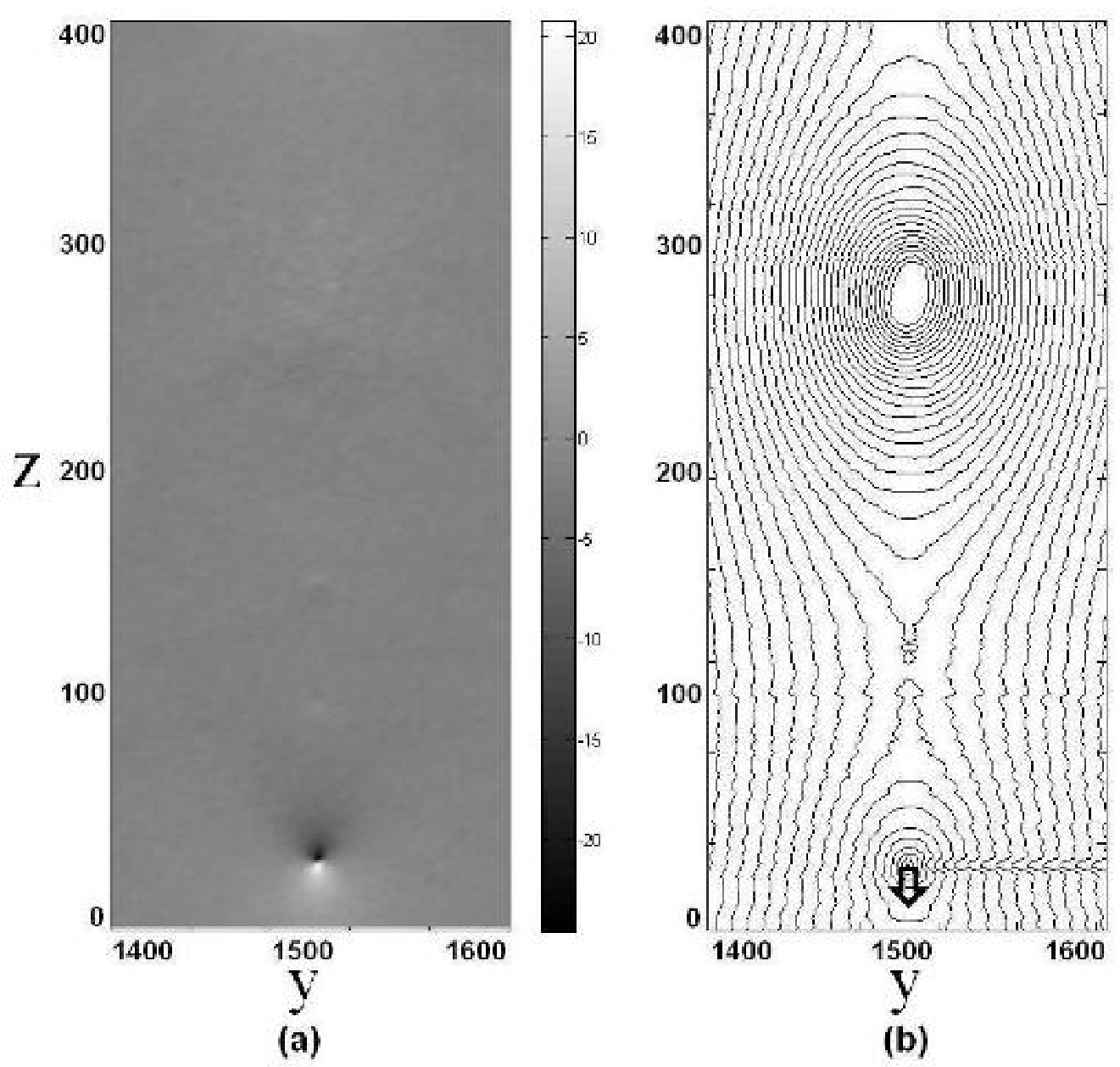}
\end{center}
\caption{ Electric (a) and magnetic (b) fields in the current sheet
at $t=9820$} \label{eb}
\end{figure}

The black and cyan particles are already outside the X-point
region; they were accelerated in this region earlier and now
gain energy by compression. One sees in Fig.\ref{chi}b that the
energy of the cyan particle varies in oscillatory manner; at
the later stage, the green particle exhibits the same
oscillatory behavior. In order to explain such oscillations,
the electric and magnetic fields at the same time $t=8260$ as
Fig.\ref{chi}a are shown in Fig.~\ref{eb}a and b, respectively.
One sees a small magnetic island in the region where the cyan
particle is located in Fig.\ref{chi}a. The island rapidly moves
away from the X-point region as shown in Fig.~\ref{eb}b by
arrows and eventually merges into the large island at the upper
half of the current sheet.
Such islands are formed occasionally
when the magnetic field lines are teared in the X-point. The
electric field within such a rapidly moving island reverses
sign according to $\bf{E}=\bf{v} \times \bf{B}$, being positive
in front and negative in back of the island motion, as one can
see in Fig.~\ref{eb}a. Note that within the island, the
electric field remains smaller than the magnetic field (even
though the magnitude of the electric field is significantly
larger within the island than near the X-point) therefore the
particles are not accelerated there. Low energy particles just
drift together with the island because their Larmor radius is
less than the size of the island however particles with the
Larmor radius comparable with the size of the island experience
oscillations.

When a small magnetic island appears, an additional X-point
arises; in Fig.~\ref{eb}b it is at the top of the box (recall
that the box is periodic in the $z$-direction). An additional
low-$\chi$ region arises there (at the top of Fig.~\ref{chi}a)
so that an additional acceleration cite appears. In
Fig.~\ref{chi}a, the red particle moves down so that it is
going to enter this region from above; Fig.~\ref{chi}b shows
that this particle is indeed accelerated soon.

\begin{figure}[h]
\begin{center}
\includegraphics[width=.5\columnwidth]{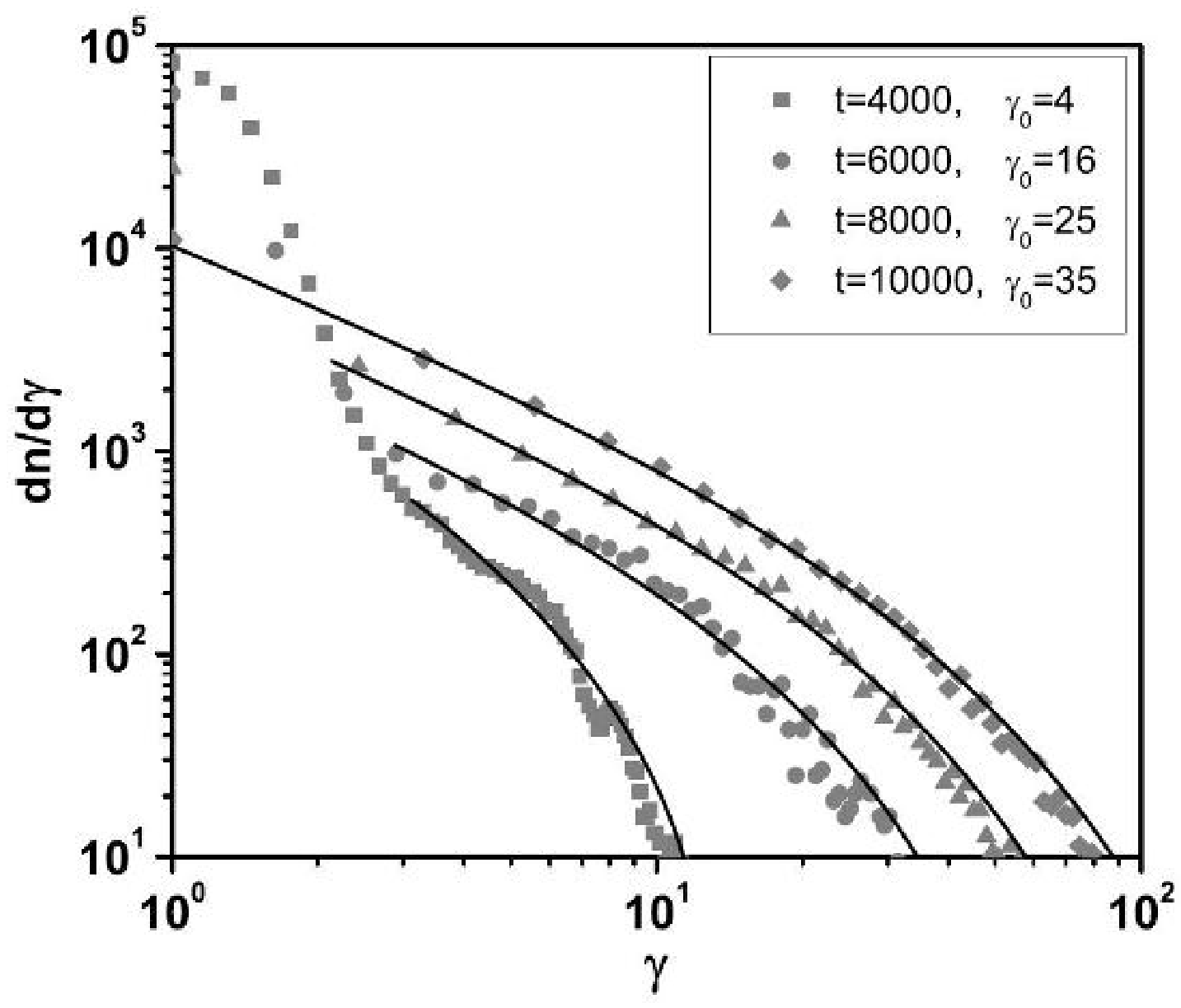}
\end{center}
\caption{Evolution of the total particle energy spectrum. The dotted
curves represent the fit (\ref{fit}). The spectrum
curves are presented for the time $t =$ $4000$, $6000$,
$8000$, $10000$ and fit $\gamma_0 =$ $4$, $16$, $25$, $35$,
correspondingly} \label{spectrum}
\end{figure}

Fig.\ref{spectrum} shows evolution of the overall energy spectrum.
One sees that the range of particles energies gradually expands
and the high-energy tail appears as plasma is compressed.  The
resulted energy spectrum is well approximated by the expression
proposed by Larrabee et al. (2003):
\begin{equation}
\frac{dn}{d\gamma}\propto\frac{1}{\gamma}\exp{\left(-\frac{\gamma}{\gamma_0}\right)}.
\label{fit}
\end{equation}
They showed that this formula fits the energy spectrum in the
vicinity of the X-point. In our simulations, the overall
particle distribution is well fitted  at high energies by
Eq.(\ref{fit}). The reason is that the high energy tail is
formed only by particles preaccelerated in the X-point. Outside
the X-point, they gain energy predominantly by compression. In
this case the final particle energy is proportional to the
initial energy so that the particle energy spectrum formed near
the X-point just shifts towards larger energies, the spectral
shape remaining unaltered.

\begin{figure}[h]
\begin{center}
\includegraphics[width=.5\columnwidth]{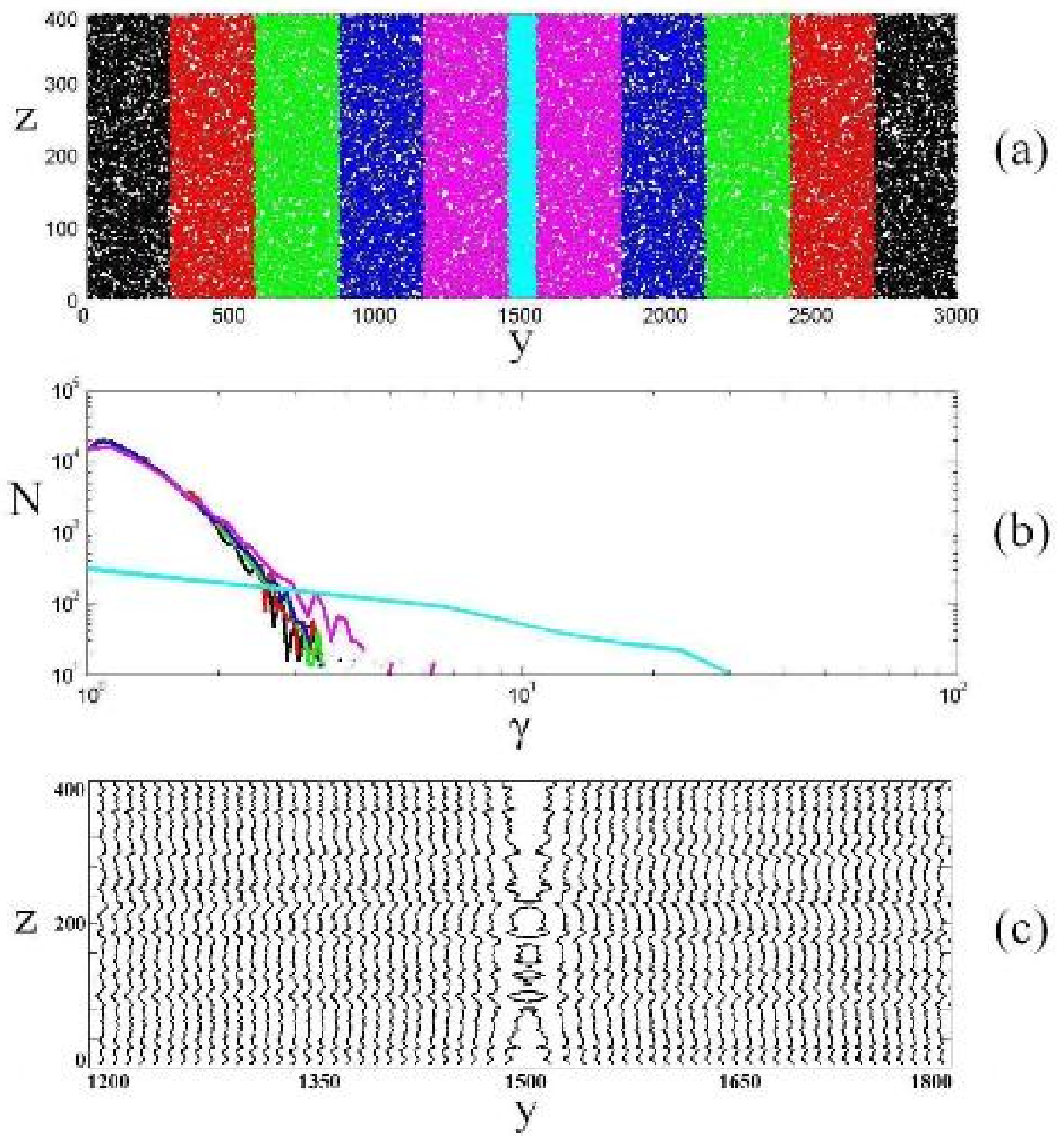}
\end{center}
\caption{The uncompressed configuration at $t=9990$: (a) particle spatial distribution, (b) energy spectrum for each of
the colored stripes and (c) magnetic field lines in the current sheet.} \label{uncompressed}
\end{figure}

Finally, in order to show that the above magnetic reconnection
is driven solely by the compression, the same simulation was
performed for the uncompressed case with constant boundary
currents on both sides. The timescale of the simulation was the
same as in the compressed case, $t=9990$, which corresponds to
$\sim 3.5$ light travel times from the boundary to the center
of the box (note that in the lobe the plasma is magnetically dominated therefore Alfv\'{e}n's velocity is close to the speed of light). Fig.~\ref{uncompressed} shows the particles final
(at $t=9990$) spatial distribution, corresponding energy
spectra and magnetic field. As one can see compared to
Figs.~\ref{motion}a and, \ref{gamma_regions}a, practically
nothing has changed as compared with the initial state; the
particles remain at their initial locations along the
$y$-direction (see Fig.~\ref{uncompressed}a) and the energy
spectra in each stripe have no considerable changes since the
initial state (see Fig.~\ref{uncompressed}b).
The small
magnetic field islands observed in the current sheet midplane
(Fig.~\ref{uncompressed}c) develop already at the early stages
of simulation whereupon nothing changes. The magnetic energy in
the box also remains constant, to within the error of
computation, throughout the simulation. These results clearly
show that the initial configuration is stable, which means that
the magnetic reconnection and the resulted particles
acceleration discussed in present paper occur exclusively due
to external compression.

\section{Conclusions}
We performed 2.5D PIC simulations of the particle acceleration
in the course of the compressionally driven reconnection in
relativistic electron-positron plasma. Since the plasma is
strongly compressed at the front of a relativistic shock, we
consider our simulations as a simple model for the particle
acceleration at the termination shock in the striped pulsar
wind. We found that the X-points play crucial role even though
a relatively small fraction of the total energy is released
there. Particles preaccelerated in the vicinity of the X-point
take a significant fraction of the released energy because in
the compressing medium, particles with larger Larmor radii gain
more energy. This result could be of more general nature
because for particles with larger Larmor radii, the frozen-in
condition is violated easier therefore one can expect that in
various situations, particles preaccelerated in the vicinity of
X-points could take a good fraction of the dissipated magnetic
energy.

In our simulations, the fraction of the particles initially
confined in the current sheet is small so that in the final state,
the energy spectrum of the particles in the sheet is independent
of the initial condition. This spectrum is found to be
approximately $dn/d\gamma \propto \gamma^{-1}$ for
$\gamma<\gamma_0$, while is rapidly decreasing for
$\gamma>\gamma_0$. The maximal energy $\gamma_0$ grows in the
course of reconnection. The electrons with the spectrum
$dn/d\gamma \propto \gamma^{-1}$ emit synchrotron radiation with
the flat spectrum. Therefore our results support the idea that
flat radio spectra of plerions may be attributed to the particle
acceleration at the termination shock in the striped pulsar wind
(Lyubarsky 2003).

Note that the very occurrence of X-points was questioned by
Zenitani \& Hoshino (2005, 2007) who showed that in the
electron-positron plasma, the tearing instability, which is
thought to be responsible for formation of X-points, develops
slower than the drift kink instability, which just shifts straight
magnetic field line tubes making the current sheet folded. Their
PIC simulations show that the field dissipation due to the
drift-kink instability does not result in non-thermal particle
acceleration, the plasma is just being heated. The reason is that
in this case, the field line tubes remain straight so that all
particles gain energy with the same rate. However, the drift kink
instability is suppressed in the presence of the current-aligned
magnetic field (the so called "guide field"); then the magnetic
reconnection occurs in X-points and produces a lot of nonthermal
particles (Zenitani \& Hoshino 2008). Therefore one can believe that the
mechanism identified in this work remains relevant at quite
general conditions.

This work was supported by the German-Israeli foundation for
scientific research and development under the grant
I-804-218.7/2003.


\begin{thebibliography}{99}

\bibitem[Birdsall \& Langdon 1985]{Birdsall85}
Birdsall, C.K., \& Langdon, A.B. 1985, Plasma Physics via Computer Simulation, (New-York: McGraw-Hill)

\bibitem[Birk et al. 2001]{Birk01}
Birk, G.T., Crusius-W\"{a}tzel, A.R., \& Lesch, H. 2001, \apj,
{\bf 559}, 1, 96
%Hard Radio Spectra from Reconnection Regions in Galactic Nuclei

\bibitem[Buneman 1993]{Buneman93}
Buneman, O. 1993, in Computer
Space Plasma Physics: Simulation Techniques and Software, ed. H. Matsumoto \& Y. Omura
(Tokyo: Terra Scientific), 67

\bibitem[Cargill 2001]{Cargill01}
Cargill, P.J. 2001, Adv. Space Res., {\bf 26}, 1759
%Theories of heating and particle acceleration in the solar corona

\bibitem[Drake et al. 2005]{Drake05}
Drake, J. F., Shay, M. A., Thongthai, W., \& Swisdak, M. 2005,
\prl, {\bf 94}, 095001
%Production of Energetic Electrons during Magnetic Reconnection

\bibitem[Drenkhahn 2002]{Drenkhahn02}
Drenkhahn, G. 2002, \aap, {\bf 387}, 714
%Acceleration of GRB outflows by Poynting flux dissipation

\bibitem[Drenkhahn \& Spruit 2002]{Drenkhahn-Spruit02}
Drenkhahn, G., \& Spruit, H.C. 2002, \aap, {\bf 391}, 1141
%Efficient acceleration and radiation in Poynting flux powered GRB outflows

\bibitem[Jaroschek et al. 2004]{Jaroschek04}
Jaroschek, C.H., Lesch, H., \& Treumann, R.A. 2004, \apj, {\bf 605}, 1, L9
%Relativistic Kinetic Reconnection as the Possible Source Mechanism for High Variability and Flat Spectra in Extragalactic Radio Sources

\bibitem[Karlicky 2007]{Karlicky07}
Karlicky, M. 2007, arXiv:0709.0572
%Separation of accelerated electrons and positrons in the relativistic reconnection

\bibitem[Kirk 2004]{Kirk04}
Kirk, J.G. 2004, \prl, {\bf 92}, 18, 181101
%Particle Acceleration in Relativistic Current Sheets

\bibitem[Kirk et al. 2007]{Kirk07}
Kirk, J.G., Lyubarsky, Y., \& Petri, J. 2007, arXiv:astro-ph/0703116
%The theory of pulsar winds and nebulae

\bibitem[Larrabee et al. 2003]{Larrabee03}
Larrabee, D.A., Lovelace, R.V.E., \& Romanova, M. M. 2003, \apj, {\bf 586}, 1, 72
%Lepton Acceleration by Relativistic Collisionless Magnetic Reconnection

\bibitem[Lindman 1975]{Lindman75}
Lindman, E.C. 1975, J. Comput. Phys. {\bf 18}, 66

\bibitem[Lyubarsky 2003]{Lyubarsky03}
Lyubarsky, Y. 2003, \mnras, {\bf 345}, 1, 153
%The termination shock in a striped pulsar wind

\bibitem[Lyubarsky 2005]{Lyubarsky05}
Lyubarsky, Y. 2005, Adv. Space Res., {\bf 35}, 1112
%The termination shock in a striped pulsar wind

\bibitem[Lyutikov 2003]{Lyutikov03}
Lyutikov, M. 2003, \mnras, {\bf 346}, 2, 540
%Explosive reconnection in magnetars

\bibitem[di Matteo 1998]{diMatteo98}
di Matteo, T. 1998, \mnras, {\bf 299}, 1, L15
%Magnetic reconnection: flares and coronal heating in active galactic nuclei

\bibitem[P\'{e}tri \& Lyubarski 2007]{Petri07}
P\'{e}tri, J., \& Lyubarsky, Y. 2007, \aap, {\bf 473}, 3, 683
%Magnetic reconnection at the termination shock in a striped pulsar wind

\bibitem[Pritchett 2006]{Pritchett06}
Pritchett, P. L. 2006, \grl, {\bf 33}, 13, L13104
%Relativistic electron production during driven magnetic reconnection

\bibitem[Romanova \& Lovelace 1992]{Romanova92}
Romanova, M. M., \& Lovelace, R.V.E. 1992, \aap, {\bf 262}, 1, 26
%Magnetic field, reconnection, and particle acceleration in extragalactic jets

\bibitem[Thompson 2006]{Thompson06}
Thompson, C. 2006, \apj, {\bf 651}, 1, 333

\bibitem[Thompson \& Duncan 1995]{Thompson95}
Thompson, C., \& Duncan, R. C. 1995, \mnras, {\bf 275}, 255
%The soft gamma repeaters as very strongly magnetized neutron stars - I. Radiative mechanism for outbursts

\bibitem[Yee 1966]{Yee66}
Yee, K.S. 1966, IEEE Trans. Antennas Propagat. {\bf 14}, 302
%Numerical Solution of Initial Boundary Value Problems Involving Maxwell's Equations in Isotropic Media

\bibitem[Zenitani \& Hoshino 2001]{Zenitani01}
Zenitani, S., \& Hoshino, M. 2001, \apj, {\bf 562}, 1, L63
%The Generation of Nonthermal Particles in the Relativistic Magnetic Reconnection of Pair Plasmas

\bibitem[Zenitani \& Hoshino 2005]{Zenitani05}
Zenitani, S., \& Hoshino, M. 2005, \apj, {\bf 618}, 2, L111
%Relativistic Particle Acceleration in a Folded Current Sheet

%\bibitem[Zenitani \& Hoshino 2005]{Zenitani05b}
%Zenitani, S., \& Hoshino, M. 2005, \prl, {\bf 95}, 9, 095001
%Three-Dimensional Evolution of a Relativistic Current Sheet: Triggering of Magnetic Reconnection by the Guide Field

\bibitem[Zenitani \& Hoshino 2007]{Zenitani07}
Zenitani, S., \& Hoshino, M. 2007, \apj, {\bf 670}, 1, 702
%Particle Acceleration and Magnetic Dissipation in Relativistic Current Sheet of Pair Plasmas

\bibitem[Zenitani \& Hoshino 2008]{Zenitani08}
Zenitani, S., \& Hoshino, M. 2008, \apj, {\bf 677}, 1, 530
%The Role of the Guide Field in Relativistic Pair Plasma Reconnection

\end{thebibliography}
\end{document}